\documentclass[preprint]{revtex4}
\usepackage{enumerate}
\usepackage{graphicx}
\usepackage{dcolumn}
\usepackage{bm}
\usepackage{amsmath}
\usepackage{amsxtra}
\usepackage{amstext}
\usepackage{amssymb}
\usepackage{latexsym}

\begin{document}

\title{Comments on "Maximum or minimum entropy production ? How to select a necessary criterion of stability for a dissipative fluid or plasma"}

\author{
Giorgio SONNINO$^{1*}$, Mustapha TLIDI$^{1}$, and Jarah EVSLIN$^{2}$\\
\vskip0.4truecm
$^1$Universit{\'e} Libre de Bruxelles (U.L.B.), Department of Theoretical Physics and Mathematics, Campus de la Plaine C.P. 231 - Bvd du Triomphe, 1050 Brussels - Belgium\\
\vskip0.1truecm
$^{2}$ Theoretical Physics Center for Science Facilities Institute of High Energy Physics, Chinese Academy of Sciences, YuQuanLu 19B, 100049 Beijing - China}

\begin{abstract}
In a recent paper, Phys. Rev E {\bf 81}, 041137 (2010), the author attempts to derive ten necessary conditions for stability of dissipative fluids and plasmas. Assuming the validity of the local equilibrium principle, these criteria have been obtained solely from the first and the second laws of thermodynamics. The Onsager reciprocity relations have not been invoked and author's results are supposed to be valid independently of the choice of the boundary conditions. In the present paper, in agreement with the general theory established by Glansdorff-Prigogine in 1954 and 1970, we shall show that there is no variational principle expressing the necessary conditions for stability of dissipative systems involving convective effects when the system is out of the Onsager region. In particular, we shall prove that the basic equations constituting the starting point of the analysis of the author, attempting to derive ten necessary conditions for the stability involving magneto-hydrodynamical effects, are incorrect and in contradiction with the laws of the thermodynamics of irreversible processes.

\vskip1truecm
\noindent *Email: gsonnino@ulb.ac.be

\end{abstract}

\maketitle


\section{Introduction}
\vskip 0.2truecm
\noindent In the recently published papers \cite{divita1}, \cite{divita2} and \cite{divita3}, the author attempts to derive ten necessary stability criteria for dissipative systems involving magneto-convective effects. These criteria take the form of - or they are derived from - variational principles and they have been obtained by invoking no Onsager symmetry and no detailed model for heat production and transport. By integrations of the inequalitiy expressing the stability condition, and the balance equations for mass and energy, the author derived a set of constraints on the evolution of smooth perturbations relaxing towards the steady-state. These constraints take the form of inequalities involving the total time derivative of quantities such as the volume of the system and the entropy produced by heating processes. Each inequality takes the form $d_tA\leq d_tB+d_tC$, where $d_t$ stands for the total (i.e., substantial) time derivative and $A$, $B$ and $C$ are volume integrals.

\noindent However, these results are manifestly in contradiction with the laws of the thermodynamics of irreversible processes. The mistake in the work published in Refs~\cite{divita1}, \cite{divita2}, is a result of a misinterpretation of Eqs~(1.5) and (1.4), reported in Ref. ~\cite{divita1} and Ref. ~\cite{divita2}, respectively. Indeed, these equations are valid only for dissipative systems in absence of convective effects \cite{glansdorff}, \cite{glansdorff1}. We would like to clarify that, contrary to the claims of Ref.~\cite{divita1}, Eqs~(1.5) and (1.4) do not correspond to the expression of the Universal Criterion of Evolution (UCE) demonstrated by Glansdorff and Prigogine in 1954. Eqs~(1.5) and (1.4), are derived from the Gibbs expression, written in local form, in conjunction with the local equilibrium stability. Hence, these equations are a direct consequence only of the second principle of thermodynamics. In the case of pure dissipative processes, the time derivative appearing in these equations should be understood as the total time derivative (in the {\it substantial} sense). However, for inhomogeneous systems in presence of convective effects, Eqs~(1.5) and (1.4) should be re-written by involving only the {\it partial} time derivatives, $\partial_t$, and by no means in terms of total time derivatives. In addition, in the presence of convective effects, the inequality for the necessary condition of stability should include the extra term $-T^{-1}(\partial_t {\bf{\rm v}})^2$ \cite{glansdorff2}, where $T$ and ${\bf{\rm v}}$ are the temperature and the velocity of the matter, respectively. This term has been omitted in Eqs~(1.5) and (1.4) of Refs~\cite{divita1} and \cite{divita2}. In addition, if ${\bf q}$ indicates the heat current associated with the heat balance equation of the form
\begin{equation}\label{i1}
\rho c_v\frac{\partial T}{\partial t}=-\nabla\cdot {\bf q}
\end{equation}
\noindent
\noindent with $\rho$ and $c_v$ denoting respectively the mass density and the specific heat at constant volume, it becomes obvious that Eq.~(2.6) in Ref.~\cite{divita1}, cannot be correct. Eq.~(\ref{i1}) leads to the inequality
\begin{equation}\label{i2}
\int_\Omega (\nabla\cdot {\bf q})\frac{\partial}{\partial t}\Bigl(\frac{1}{T}\Bigr)\ d{\rm V}\geq 0
\end{equation}
\noindent where $\Omega$ is the volume occupied by the system. However, expression~(2.6), deduced in Ref.~\cite{divita2}, exhibits the reverse inequality. The author, attempting to derive the ten necessary stability criteria, relied heavily upon the inequality (2.6), which is manifestly in contradiction with the second law of thermodynamics. The ten stability criteria, taking the form of inequalities of the type $d_tA\leq d_tB+d_tC$ (see Eqs~(3.1), (3.2) and (3.3) in Ref.~\cite{divita1}), have been deduced from Eqs~(1.5) and from the inequality (2.6). Hence, they are also incorrect. The application of the the ten stability criteria to the thermonuclear reactor IGNITOR is also incorrect as the discussion revolves around the (incorrect) Eq.~(3.1) in Ref.~\cite{divita1}.

\noindent The present comments are organized as follows. The expression for the second order differential of entropy, obtained following the original demonstration of Glansdorff-Prigogine is illustrated in Section (\ref{thermo}). The link between the Le Ch$\hat{\rm a}$telier-Braun principle and the local stability condition is also shown in this section. The main conclusion of our analysis can be found in Section (\ref{cs}).
\vskip 0.2truecm
\section{Second Order Differential of Entropy}\label{thermo}
\vskip 0.2truecm
\noindent Let us introduce the Glansdorff-Prigogine increment $\delta$ \footnote{Notice that in the Glansdorff-Prigogine formalism, symbol $\delta$ expresses the variation of a physical quantity with respect to time or to a particular spacial direction $\xi$ (with $\xi=x$ or $\xi=y$ or $\xi=z$). As it is clearly pointed out by P. Glansdorff-I. Prigogine in Ref.~\cite{glansdorff3}, these variations must be performed one at a time to avoid making mistakes when one computes the second order differential, $\delta^2$, of a physical quantity $\mathcal A$. Hence, in the P. Glansdorff-I. Prigogine expressions for $\delta^2{\mathcal A}$, we are allowed to replace $\delta$ {\it either} with $\partial_t$, {\it or} with $\partial_\xi$. It is evident, however, that in the case of a combination of variations (like for example the variation of a physical quantity ${\mathcal A}$ along a line ${\bf x}(t)$, which is made according to the rule $d_t{\mathcal A}=\partial_t{\mathcal A}+d_t{\bf x}\cdot\nabla {\mathcal A}$), the correct expressions involving the total derivatives $d_t$, are by no means obtained simply by replacing, in the P. Glansdorff-I. Prigogine expressions for $\delta^2{\mathcal A}$, the increments $\delta$ with $d_t$. The mistake in the work published in Refs \cite{divita1} and \cite{divita2} is a result of a misinterpretation of the commented-on author of symbol $\delta$ introduced by P. Glansdorff-I. Prigogine.}. Our aim is to derive a set of relations coming from the second order quantity $\delta^2 s$. The Gibbs relation for $\delta s$ reads:
\begin{equation}\label{s1}
T\delta s=\delta u +p\delta v-\sum_i\mu_i\delta N_i
\end{equation}
\noindent where $s$ indicates the total entropy of the system per unit mass and $u$ and $N_i$ denote the energy density per unit mass and the mass fractions, respectively. $p$, $v$ and $\mu_i$ are the pressure, the specific volume ($v=\rho^{-1}$, with $\rho$ indicating the total density), and the chemical potentials per unit mass, respectively. By simple differentiation of Eq.~(\ref{s1}) and by expressing $\delta\mu_i$ in terms of the variables $T$, $p$ and $N_i$, we obtain the second order quantity $\delta^2 s$ \cite{glansdorff3}
\begin{align}\label{s2}
&\delta^2 s=-\frac{1}{T}\Bigl[\frac{c_v}{T}(\delta T)^2+\frac{\rho}{\chi}(\delta v)^2_{N_i}+\sum_{ij}\Bigl(\frac{\partial\mu_i}{\partial N_j}\Bigr)_{T,p,(N_i)}\delta N_i\ \delta N_j\Bigr]\qquad\quad {\rm where}\\
&(\delta v)_{N_i}\equiv \Bigl(\frac{\partial v}{\partial T}\Bigr)_{p,N_i}\delta T+\Bigl(\frac{\partial v}{\partial p}\Bigr)_{T,N_i}\delta p\nonumber
\end{align}
\noindent Here, $\chi$ is the thermal dilatation coefficient at constant pressure. The index $(N_i)$ means that all mass fractions, except $N_i$, are maintained constant. An identical calculation, in which the operator $\delta$ is replaced by the time partial derivative $\partial_t$, yields the equality \cite{glansdorff3}
\begin{align}\label{s3}
&\partial_tT^{-1}\partial_t(\rho u)-\rho\sum_i\partial_t(\mu_i T^{-1})\partial_t N_i+\rho T^{-1}\partial_tp\ \partial_tv-h\partial_tT^{-1}\partial_t\rho=\\
&\qquad\quad-\rho T^{-2}\Bigl(\frac{\partial u}{\partial T}\Bigr)_{v,N_i}(\partial_t T)^2+\rho T^{-1}(\frac{\partial v}{\partial p}\Bigr)^{-1}_{T,N_i}\Bigl[\Bigl(\frac{\partial v}{\partial T}\Bigr)_{p,N_i}\partial_tT+\Bigl(\frac{\partial v}{\partial p}\Bigr)_{T,N_i}\partial_t p\Bigr]^2
\nonumber\\
&\qquad\quad -\rho T^{-1}\sum_{ij}\Bigl(\frac{\partial\mu_i}{\partial N_j}\Bigr)_{T,p,(N_i)}\partial_t N_i\ \partial_t N_j
\nonumber
\end{align}
\noindent where $h$ stands for the enthalpy per unit mass: $h=u+p\rho^{-1}$. It should be kept in mind that in Eq.~(\ref{s3}), the independent variables ($u,v,N_i$) characterize the local state of a {\it dissipative system} i.e., the convective effects are neglected. According to the thermodynamic stability theory, a state is defined to be {\it stable} if no evolution starting from the unperturbed state can satisfy the requirements of the second law. In the presence of hydrodynamic effects,the generalized sufficient condition of local stability, valid for convective as well as for dissipative processes, takes the form \cite{glansdorff2}, \cite{glansdorff3}
\begin{equation}\label{s4}
\partial_t\delta^2 z\geq0\quad; \quad\delta^2z<0
\end{equation}
\noindent where 
\begin{equation}\label{s5}
\delta^2z\equiv\delta^2s-T^{-1}(\delta{\rm v})^2=\delta T^{-1}\delta u+\delta(pT^{-1})\delta v-\sum_i\delta(\mu_i T^{-1})\delta N_i-T^{-1}(\delta{\rm v})^2 < 0
\end{equation}
\noindent Hence, in case of time-dependent convection processes, a supplementary contribution equal to $-T^{-1}(\delta {\rm\bf v})^2$ should be added to the second variation of entropy. Of course, in presence of hydrodynamic effects the fields, like the temperature, are linked to the velocity through the balance equations for mass, energy and momentum. This extra term has been omitted in Eq.~(1.5) and in Eq.~(1.4) of Refs.~\cite{divita1} and \cite{divita2} respectively. Notice that, as clearly explained by P. Glansdorff and I. Prigogine (see Ref.~\cite{glansdorff3}, chapter VI, \S  7), the terms $-T^{-1}(\delta {\bf v})^2 $ is of the same order as the other contributions and then, by no means can this term be neglected. We conclude this comments by recalling that the {\it Le Ch$\hat a$telier-Braun Principle} affirms that {\it as a result of a variation in one of the factors governing the thermodynamic equilibrium of a system, the system tends to adjust to a new equilibrium state counteracting the imposed change}. It has been proven that, if using $\delta^2z$ as the Lyapunov function the system is locally stable then the Le Ch$\hat a$telier-Braun Principle is automatically satisfied \cite{glansdorff}. Using Eq.~(\ref{s2}) with the additional term $-T^{-1}(\delta {\rm\bf v})^2$, the second global stability condition reads
\begin{equation}\label{s6}
-\frac{1}{T}\Bigl[\frac{c_v}{T}(\partial_t T)^2+\frac{\rho}{\chi}(\partial_t v)^2_{N_i}+\sum_{ij}\Bigl(\frac{\partial\mu_i}{\partial N_j}\Bigr)_{T,p,(N_i)}\partial_t N_i\ \partial_t N_j+\sum_i(\partial_tv_i)^2\Bigr]\leq0
\end{equation}
\noindent where ${\rm v}_i$ denotes the $ith$ component of the velocity of matter. Let us now consider systems subject to {\it time-independent boundary conditions} and, in particular, to the additional boundary conditions $[\partial_t({\rm v}_i)]_\Sigma=0$. Using these boundary conditions, and by taking into account the balance equations for mass, energy and momentum, we derive the celebrated {\it Universal Criterion of Evolution} (UCE) by integrating Eq.~(\ref{s5}) over the volume occupied by the system \cite{glansdorff2}
\begin{equation}\label{s7}
\int_\Omega\sum_\kappa J_k\partial_tX_\kappa\ d{\rm V}\leq0
\end{equation}
\noindent where $J_\kappa$ and $X_\kappa$ indicate the {\it thermodynamic flows} and the {\it thermodynamic forces}, respectively. As an example of application, we may re-consider the stability of heat conduction in a solid where the temperature is the unique variable. For time-independent boundary conditions, it is easy to check that, by applying the divergence theorem, we obtain again inequality ~(\ref{i2}) \cite{groot}.
\vskip 0.2truecm
\section{Conclusions}\label{cs}
\vskip 0.2truecm
The mistake in the work published in Refs~\cite{divita1} and \cite{divita2} is the result of a wrong interpretation of the equation for the second order differential of entropy. In the case of dissipative fluids and plasmas
\begin{description}
\item {\it {\bf A.} By no means can the increment $\delta$ be replaced by the substantial derivative $d_t$ in the equation for $\delta^2s$};
\item {\it {\bf B.} The inequality for the necessary condition for stability should include the additional term $-T^{-1}(\delta{\bf{\rm v}})^2$}.
\end{description}
\noindent In the papers \cite{divita1} and \cite{divita2} we found the (fundamental) mistake reported in {\bf A.} and the extra term, mentioned in {\bf B.}, is totally absent. In addition Eq.~(2.6), deduced by the author in Ref.~\cite{divita2}, and largely used in his demonstration, is in contradiction with the general expression of the UCE. Since Eqs~(1.5) and (2.6), in Ref.~\cite{divita1}, are {\it the starting point} of the author's attempt to derive ten necessary conditions for the stability of non-equilibrium magneto-hydrodynamic systems, the final inequalities (3.1), (3.2) and (3.3) (in Ref.~\cite{divita1}), as well as the derived criteria for stability, are also incorrect.

\noindent The paper cited in Ref.~\cite{divita3} is an application of the (incorrect) variational results published in Refs~\cite{divita1} and \cite{divita2} to the thermonuclear reactor IGNITOR. Indeed, the discussion revolves around Eq.(1) reported in Ref.~\cite{divita3} [which corresponds to Eq. (3.1) in Ref.~\cite{divita1}]. However, as stated above, this equation is incorrect and, as a consequence, the conclusions are also incorrect. Anyway, calculations for IGNITOR-plasmas, in the collisional transport regimes, have been performed recently \cite{sonnino}. The results are not in contradiction with the general statements reported in Ref.~\cite{bombarda} and with the UCE. 

\noindent We conclude with a quotation from the original work of Glansdorff and Prigogine on the Universal Criterion of Evolution.

\noindent "{\it The sign, which corresponds to the exact differential of the total change of the entropy production, is by no means prescribed by the Universal Criterion of Evolution}".

\section{Acknowledgments}
We are grateful to Prof. R. Lefever, of the Universit{\' e} Libre de Bruxelles, for his suggestions and the scientific discussions held in this subject.  One of us (G.S.), is very grateful to Prof. M.Malek Mansour, of the Universit{\'e} Libre de Bruxelles, for his scientific suggestions. J.E. is supported by the Chinese Academy of Sciences Fellowship for Young International Scientists grant number 2010Y2JA01. M.T. is a Research Associate with the Fonds de la Recherche Scientifique F.R.S.-FNRS, Belgium.




\begin{thebibliography}{alpha}

\bibitem{divita1} A. Di Vita, Phys. Rev E {\bf 81}, 041137 (2010).

\bibitem{divita2} A. Di Vita, Proc. R. Soc. Lond. A {\bf 458}, 21 (2002).

\bibitem{divita3} A. Di Vita, Nucl. Fusion {\bf 50}, 115001 (2011).

\bibitem{glansdorff} P. Glansdorff and I. Prigogine {\it Physica}, {\bf 20}, 773 (1954).

\noindent P. Glansdorff and I. Prigogine {\it Physica}, {\bf 30}, 351 (1964).

\bibitem{glansdorff1} \noindent P. Glansdorff and I. Prigogine {\it Physica}, {\bf 31}, 1242 (1965).

\bibitem{glansdorff2} P. Glansdorff and I. Prigogine {\it Physica}, {\bf 46}, 344 (1970).

\bibitem{glansdorff3} P. Glansdorff and I. Prigogine (1971), {\it Thermodynamic Theory of Structure, Stability and Fluctuations}, Wiley-Interscience, London-New york-Sydney-Toronto.

\bibitem{groot} S.R. De Groot and P. Mazur (1971), {\it Non-Equilibrium Thermodynamics} Dover Publications, Inc., New York.

\bibitem{sonnino} G. Sonnino {\it et al.}, {\it Losses and Distribution Functions for Collisional IGNITOR-Plasmas}, to be sent to {\it Nucl. Fusion} (2012).

\bibitem{bombarda} F. Bombarda {\it et al.}, Braz. J. Phys. {\bf 34}, 1786 (2004).

\end{thebibliography}
\end{document}